\documentclass{IEEEtran}
\usepackage{cite}
\usepackage{amsmath,amssymb,amsfonts}
\usepackage{algorithmic}
\usepackage{graphicx}
\usepackage{textcomp}
\def\BibTeX{{\rm B\kern-.05em{\sc i\kern-.025em b}\kern-.08em
    T\kern-.1667em\lower.7ex\hbox{E}\kern-.125emX}}

\usepackage{adjustbox} 
\usepackage{booktabs}
\usepackage{multirow}
\usepackage{array}
\usepackage{colortbl}
\usepackage{bbding}
\usepackage{pifont}
\usepackage[table]{xcolor}
\usepackage{soul}

\usepackage{soul}
\sethlcolor{white} 

\newcolumntype{C}[1]{>{\centering\let\newline\\\arraybackslash\hspace{0pt}}m{#1}}
    
\begin{document}

\title{Wearable Technologies for Monitoring\\Upper Extremity Functions During Daily Life\\in Neurologically Impaired Individuals}

\author{Tommaso~Proietti$^{1}$,
        Andrea~Bandini$^{1,2}$ 
\thanks{Work supported by \#NEXTGENERATIONEU (NGEU) and funded by the Ministry of University and Research (MUR), National Recovery and Resilience Plan (NRRP) with two projects: project THE (IECS00000017) - Tuscany Health Ecosystem (DN. 1553 11.10.2022) and project MNESYS (PE0000006) – A Multiscale integrated approach to the study of the nervous system in health and disease (DN. 1553 11.10.2022); This project has been funded partly by the Bertarelli Foundation and by the CHRONOS Project funded by the Swiss National Science Foundation (grant number: 184847).\\$^1$ The Biorobotics Institute and Department of Excellence in Robotics and AI, Scuola Superiore Sant'Anna, Pisa, Italy.\\$^2$ Health Science Interdisciplinary Research Center, Scuola Superiore Sant'Anna, Pisa, Italy.}
}

\maketitle

\begin{abstract}

Neurological disorders, including stroke, spinal cord injuries, multiple sclerosis, and Parkinson's disease, generally lead to diminished upper extremity (UE) function, impacting individuals' independence and quality of life. Traditional assessments predominantly focus on standardized clinical tasks, offering limited insights into real-life UE performance. In this context, this review focuses on wearable technologies as a promising solution to monitor UE function in neurologically impaired individuals during daily life activities. Our primary objective is to categorize the different sensors, review the data collection and understand the employed data processing approaches. After screening over 1500 papers and including 21 studies, what comes to light is that the majority of them involved stroke survivors, and predominantly employed accelerometers or inertial measurement units to collect kinematics. Most analyses in these studies were performed offline, focusing on activity duration and frequency as key metrics. Although wearable technology shows potential in monitoring UE function in real-life scenarios, it also appears that a solution combining non-intrusiveness, lightweight design, detailed hand and finger movement capture, contextual information, extended recording duration, ease of use, and privacy protection remains an elusive goal. These are critical characteristics for a monitoring solution and researchers in the field should try to integrate the most in future developments. Last but not least, it stands out a growing necessity for a multimodal approach in capturing comprehensive data on UE function during real-life activities to enhance the personalization of rehabilitation strategies and ultimately improve outcomes for these individuals.

\end{abstract}


\IEEEpeerreviewmaketitle

\section{Introduction}
\label{sec:introduction}

\IEEEPARstart{T}{he} functional use of the upper extremities (UEs) is a paramount aspect of daily life for every human being, as it directly correlates with the ability to independently conduct activities of daily living (ADLs) \cite{Schambra2019, Dollar2014}. Neurological disorders such as spinal cord injuries (SCI), stroke, multiple sclerosis (MS), and Parkinson's disease (PD) have direct consequences on the ability to use the UEs, resulting in reduced independence, diminished quality of life, and limited social participation \cite{Moulaei2022, Anderson2004, Simpson2012, Nichols-Larsen2005}. Over the years, various rehabilitative approaches, including technologies like functional electrical stimulation (FES), transcutaneous spinal cord stimulation (tSCS), and exoskeletons, in conjunction with targeted physiotherapy and occupational therapy, have made notable strides in improving UE function following neurological disorders \cite{Popovic2011, Marquez-Chin2020, Inanici2021, deFreitas2021, Proietti2022}.

Despite significant progress in the field of rehabilitation and assistive technologies, a substantial challenge persists. Traditional rehabilitation assessments and monitoring technologies predominantly concentrate on evaluating the capacity domain of UE use, as defined by the International Classification of Functioning, Disability, and Health (ICF) framework \cite{ICF}. This domain focuses on an individual's ability to perform standardized tasks in controlled environments. However, the heart of neurorehabilitation lies in the crucial aspect of translating the UE function improvements observed in clinical settings into real-life enhancements \cite{Waddell2017, Lang2021, Barth2021, Rand2012}. This translation is essential for enabling individuals with neurological impairments to achieve increased independence. To achieve this, it is crucial to assess the performance domain of the ICF, which gauges how individuals carry out activities in their typical daily environment \cite{Marino2007}.

To gain a comprehensive understanding of the impact of rehabilitative interventions on individuals with neurological disorders affecting UE functions and to tailor rehabilitation strategies that enhance their independence and social participation, it is imperative to bridge the gap between capacity and performance evaluations \cite{Bandini2022, Dousty2023, Tsai2023}. This is where the potential of monitoring individuals during their daily activities becomes evident. Observing and quantifying how patients with neurological impairments use their UEs outside of clinical environments can provide invaluable insights for fine-tuning rehabilitation programs, with the ultimate goal of improving their independence and quality of life. This approach is pivotal for crafting personalized rehabilitation strategies that address the specific needs and challenges faced every day by these individuals.

To achieve this goal, we stand at the crossroads of rehabilitation science, wearable technology, and data science. The advent of wearable sensor technologies presents an encouraging opportunity to track UE usage in individuals with neurological conditions as they go about their everyday lives. These technologies hold the promise of revealing a wealth of data that was previously unattainable within the limits of clinical environments. Moreover, thanks to recent developments in machine learning, this data can be efficiently processed and conveyed to clinicians, offering valuable insights into patients' real-life progress \cite{Kadambi2023}.

Within this context, this literature review focuses on the latest advancements in wearable technologies designed to monitor UE function in real-life, unconstrained situations (\textit{i.e.}, not limited to standardized tasks), and across various neurological conditions. By focusing on the pivotal aspect of monitoring UEs during the unstructured activities of daily life, we want to identify and categorize the most prevalent types of wearable sensors employed for monitoring UE use during daily life activities, describe the way these were used, and how collected data was processed. 

While similar previous reviews recognized the significance of wearable sensors for telemonitoring and telerehabilitation, they primarily focused on specific aspects (\textit{e.g.}, specific clinical conditions, specific biological joints, specific standard clinical assessments). For instance, Toh \textit{et al.} \cite{Toh2023} investigated the effectiveness of wearable technologies in home-based physical rehabilitation for stroke only, while Gopal \textit{et al.} \cite{Gopal2022} examined the use of wearables, smartphone-based, and tablet-based apps for standardized clinical assessments of hand function only in chronic neurological disorders. Other reviews delved into the transition of inertial sensors from laboratory to community settings for monitoring upper and lower limb functions in individuals with PD \cite{Sica2021}, explored wearable sensor data's role in stroke rehabilitation \cite{Kim2022}, and discussed wearable solutions in the context of MS \cite{Alexander2021}. \hl{This field of research has seen significant progress not only in healthcare, but also in other areas such as sports and athlete performance monitoring. In sports, wearable sensors are used to track and analyze the performance, health, and recovery of athletes, showcasing the broad applicability and potential of these technologies across diverse domains.} \cite{Rana2021}.

Given the increasing efforts to implement both off-the-shelf and customized solutions, our comprehensive perspective covering the existing literature aim to investigate the benefits and challenges of monitoring UE use in home settings in individuals with neurological diseases, and to recognize the importance of extending functional recovery beyond clinical settings. Moreover, we seek to offer insights into future research directions to researchers and practitioners alike.

\begin{figure}[t]
    \centering
    \includegraphics[width=.45\textwidth]{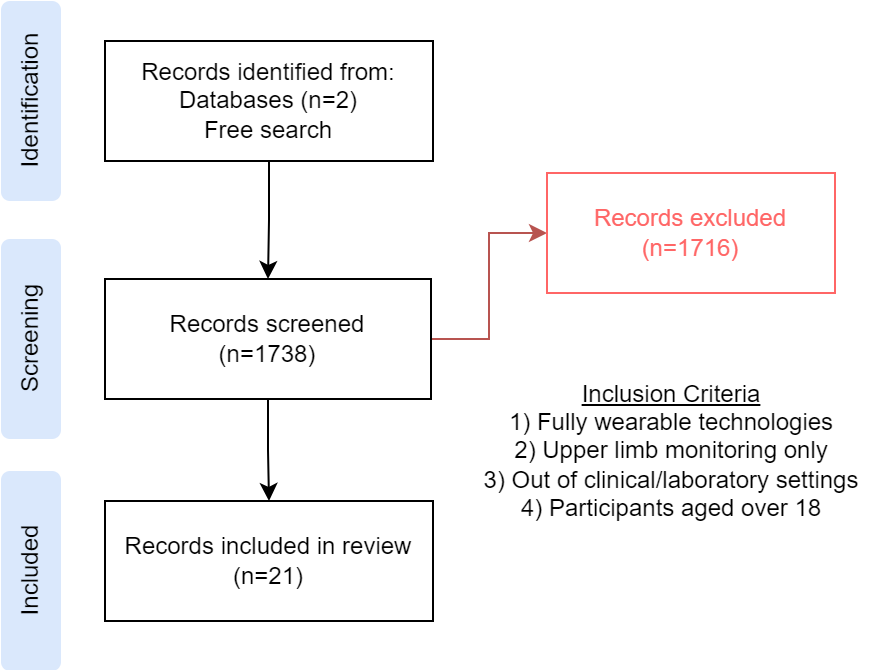}
    \caption{Review flow diagrams following PRISMA standards \cite{Haddaway2022}.}
    \label{prisma}
\end{figure}

\section{Methods}
The review was conducted by searching major scientific databases (Scopus, Google Scholar), using various combinations of keywords related to wearable technology, upper limb monitoring, and real-world applications. The search was limited to English manuscripts and journal articles only.

In the initial database search covering the period until April 2023, we retrieved a total of 1,738 titles. The first screening was performed based on the title and abstract, applying specific inclusion and exclusion criteria. Inclusion criteria comprised: 1) fully wearable technologies \hl{(e.g. armbands or sensors integrated into clothes)}, 2) upper limb monitoring \hl{(upper arms, forearms, hands)}, 3) experiments conducted outside clinical or laboratory settings \hl{(e.g. at home)}, and 4) the inclusion of adult participants aged over 18, with neurological impairments. On the contrary, we excluded studies involving healthy individuals only, non-neurological upper limb impairments (\textit{e.g.}, amputation, traumatic orthopedic conditions), papers focusing solely on lower limb monitoring, physiological signal monitoring (\textit{e.g.}, cardiovascular, brain, respiration), \hl{studies enrolling children or underage individuals,} and studies conducted solely in clinical or laboratory environments. \hl{We also excluded studies monitoring individuals via smartphone only.}

In cases of uncertainty, we reviewed the full text before making inclusion/exclusion decisions. Additionally, we included other documents found through a manual search of references from existing papers. Finally, 21 papers -- published between 2007 and April 2023 -- met the criteria for this review.

\begin{figure*}[t]
    \centering
    \includegraphics[width=.7\textwidth]{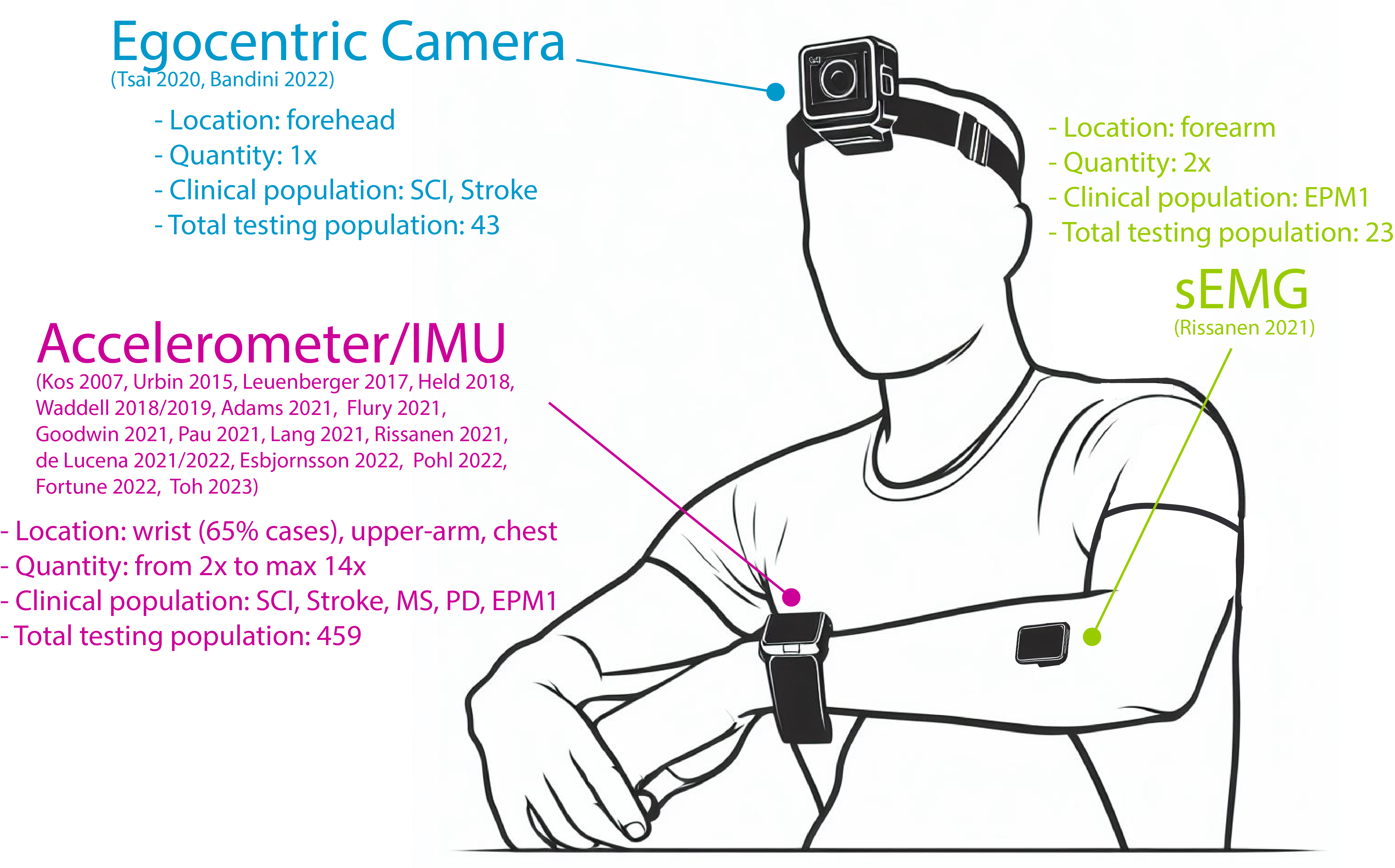}
    \caption{Overview of the main sensing technologies, body locations, and cumulative testing sizes for monitoring individuals with neurological impairments in the community. Two accelerometers or inertial measurement units (IMUs, which usually include a 3-axis accelerometer and a 3-axis gyroscope, less frequently a 3-axis magnetometer too), placed on both wrists, were the most used configuration for remote monitoring.}
    \label{sketch_man_sensors}
\end{figure*}

\section{State of the Art}\label{sec:stateoftheart}

The included studies underwent a data extraction process to identify key domains for categorizing the literature in this research field. These domains were 1) the specific neurological impairments studied for monitoring the UE functions (\textit{application}); 2) the \textit{hardware} employed for UE function monitoring ; 3) the \textit{study protocols} implemented for recording individuals with neurological impairments at home; 4) the \textit{data processing} techniques utilized to extract valuable measures to quantify UE performance from raw out-of-clinic data; 5) \textit{comparison} and analysis between clinical evaluations and at-home or community-based evaluations. In the subsequent sections, we will delve into each of these five domains. Table \ref{table_sota} and Table \ref{table_measures} detail these studies and their main characteristics.

\subsection{Application}
The distribution of cases in the reviewed studies reflects the incident rates of the corresponding neurological diseases that lead to UE impairments, which subsequently result in challenges when performing daily activities at home. Therefore, the vast majority of investigations focused on individuals with stroke (64\%). Stroke affects approximately 1 in 4 people worldwide \cite{Owolabi2022} and it is the third-leading cause of death and disability worldwide \cite{Feigin2021}. One of the most common consequences of stroke is UE impairment \cite{Hayward2019}, which makes it a significant area of research in the wearable technology domain. Following, studies examining SCI constituted 18\%, MS accounted for 9\%, PD and progressive myoclonic epilepsy type 1 (EPM1) both for 1 only out of 21, all conditions affecting less than 1\% of the global population, according to data sourced from the World Health Organization \cite{WHO1, WHO2, WHO3}. On average, the sample size was $24\pm16$ participants per study. Only two studies \cite{Lang2021, Waddell2018} enrolled more than 50 participants, while 10 out of 21 studies enrolled less than 20 individuals. Among the studies enrolling post-stroke individuals, 31\% of studies enrolled acute or sub-acute individuals (\textit{i.e.}, within 8 weeks since the event). This data is particularly interesting and in contrast with assistive technologies studies (\textit{e.g.}, robotic exoskeletons), where most of the available literature is on chronic patients \cite{Proietti2022}.

All these neurological conditions present diverse challenges to UE function. Stroke often leads to hemiparesis, spasticity, and fine motor skill loss, impairing arm, hand, and finger movements and hindering object manipulation during daily activities \cite{Coupar2012}. Given these characteristics, most of the studies monitoring stroke individuals focused on understanding the use of the impaired side compared to the healthy one during daily life. Cervical SCI, instead, results in tetraplegia, accompanied by spasticity, loss of hand motor function, and sensation \cite{Mateo2015}, while MS typically manifests as muscle weakness and dysmetria, affecting arm and hand coordination \cite{Bertoni2015}. For these conditions, the focus was more on monitoring participant independence and ability to perform any activities of daily living. Finally, in PD and EPM1, attention is focused on identifying tremors \cite{Bloem2022} and myoclonus \cite{Rissanen2021}, respectively, which can disrupt daily activities by interfering with hand interaction and object manipulation.

\subsection{Hardware}
From a hardware standpoint, the most common strategy to monitor UE functions in individuals with neurological impairments was the use of accelerometers and inertial measurement units (IMUs), the latter generally being composed of a 3-axis accelerometer, a 3-axis gyroscope, and -- depending on the sensors -- a 3-axis magnetometer (see Figure \ref{sketch_man_sensors}). While accelerometers are capable of measuring linear acceleration, used for computing activity counts and duration metrics (see Section \ref{sec:dataProcessing}), IMUs, through sensor fusion techniques \cite{Vitali2020}, can reconstruct their pose and orientation in the 3D world, enabling tracking of the UE kinematics. Both technologies are widely used due to their ability to record continuously for hours (typical runtime on batteries is around 30 days \cite{Kos2007, Urbin2015, Waddell2019, Waddell2018, Goodwin2021, Lang2021, Pau2021, Fortune2022}), with the option to store a limited amount of data on the device's internal memory (usually ranging from few MB -- e.g. \cite{Kos2007, Urbin2015, Waddell2019, Waddell2018} -- to tens of GB on modern devices \cite{deLucena2021, deLucena2022, Pohl2022, Lang2021, Pau2021, Goodwin2021}  ). They are also user-friendly, often integrated into textiles and communicating wirelessly with computers. Additionally, they are cost-effective (commercial solutions range from \$300-400 for single sensors \cite{Lang2021, Pau2021, Waddell2019, Waddell2018} to a few thousand dollars for full suits embedding 15-20 IMUs\cite{Held2018}), compact (accelerometers are often embedded in smartwatches\cite{Lang2021, Pau2021, Waddell2019, Waddell2018}, while IMUs are typically encapsulated in small boxes placed on the body through velcro-straps), and lightweight (weighing within 50g per sensor, including batteries and electronics). In the reviewed studies, accelerometers and IMUs were generally placed on the wrist (unilaterally \cite{Kos2007,deLucena2021,deLucena2022} or bilaterally \cite{Leuenberger2017, Held2018, Lang2021, Pau2021, Esbjornsson2022, Pohl2022, Waddell2019, Waddell2018, Urbin2015}), forearms \cite{Held2018, Adams2021, Flury2021, Rissanen2021}, upper-arms \cite{Held2018, Goodwin2021, Fortune2022}, and on the chest \cite{Held2018, Goodwin2021, Adams2021, Flury2021, Fortune2022}, mostly as a reference to compensate for trunk movements. 

\begin{figure*}[t]
    \centering
    \includegraphics[width=.9\textwidth]{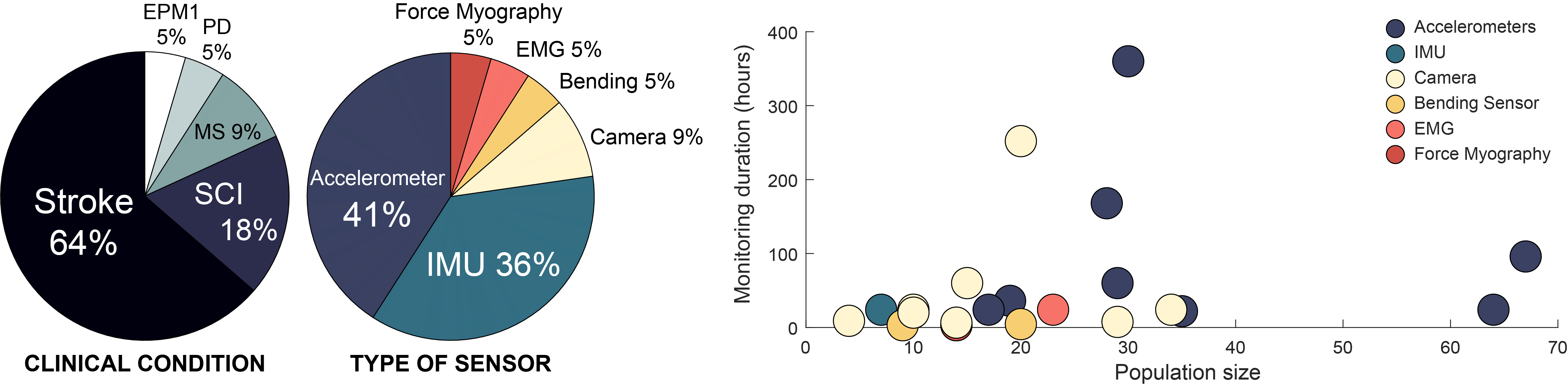}
    \caption{On the left, pie-charts with the most common clinical conditions and sensing modalities. Stroke participants wearing one or more accelerometers or IMUs was the most common scenario. On the right, a scatter plot of population size \textit{vs} monitoring duration in the 21 reviewed papers, grouped by sensor type. The majority of the studies monitored less than 30 individuals, for less than 100 hours.}
    \label{piecharts}
\end{figure*}

Only 4 out of 21 studies did not use these sensors: Bandini \textit{et al.} \cite{Bandini2022} and Tsai \textit{et al.} \cite{Tsai2020} both used an egocentric camera (Hero 4 or 5, GoPro, 120g including batteries, about \$400, approximately 2h battery runtime at 1080p and 30fps), mounted on the forehead, to record activities that involved the use of the hands during daily life; Saleh \textit{et al.} \cite{Saleh2016} manufactured conductive ink-based bending sensors integrated into the index and middle fingers of a glove to monitor hand movements (within \$100 of costs, 48h battery runtime); Yang \textit{et al.} \cite{Yang2021} used the TENZR, a commercial force myography to detect the state of the hand by monitoring the surface (\textit{i.e.}, skin) stiffness on the wrist musculo-tendonous complex. Instead, in 2 studies, accelerometers/IMUs were used in combination with other sensors. De Lucena \textit{et al.} \cite{deLucena2021, deLucena2022} embedded an IMU and four external magnetometers into a custom wrist-watch, and used this in combination with a magnetic ring on the index finger to track index and wrist movements (24h battery runtime). Rissanen \textit{et al.} \cite{Rissanen2021} used surface electromyography (sEMG) sensors placed on the forearms of the participants in combination with accelerometers. 

A few research groups designed their devices from scratch \cite{Kos2007, Saleh2016, deLucena2021, deLucena2022}, while most of the investigations were carried out using off-the-shelf technology (with classical suppliers as \textit{GoPro, ActiGraph, Xsens, APDM}). From a real-time data standpoint, only in one of the reviewed papers \cite{Esbjornsson2022} data was collected and remotely available to the research team, and only in four of the studies \cite{Saleh2016, Esbjornsson2022, deLucena2021, deLucena2022} data were available in real-time to the participants as a feedback of their performance.

Finally, it is interesting to note that most of the studies reviewed in this work did not consider comfort and wearability during their analysis. Saleh \textit{et al.} \cite{Saleh2016} stands out as the only study where the comfort of the monitoring device was assessed through a custom questionnaire administered at the end of the study. The results showed good acceptance by the participants of their custom glove. In contrast, Kos \textit{et al.} \cite{Kos2007} found that the wrist IMU was comfortable and well tolerated, while the ankle one was more visible and less appreciated. Tsai \textit{et al.} \cite{Tsai2020} reported discomfort with the forehead-mounted camera after one hour due to weight and heat, which was then confirmed by a mixed-method study conducted on the use of wearable cameras at home \cite{Bandini2021}. Yang \textit{et al.} \cite{Yang2021} mentioned that their custom force myograph was uncomfortable for two participants. All the other studies did not provide any comments or qualitative feedback on comfort and wearability, which are crucial aspects to characterize the performance of monitoring devices, particularly regarding long-term usage and user acceptance. 

\subsection{Study protocol}
From a study protocol point of view, most of the reviewed work had similar characteristics. They employed a variety of tasks to record and assess UE function, encompassing typical daily routines and ADLs that are relevant to individuals in their home environments. As an example, Saleh \textit{et al.} \cite{Saleh2016} investigated a broad range of activities, including household tasks, writing, using a remote control, dressing, grooming, working on a computer, tying shoes, and using a phone, all performed using the impaired hand. While Rissanen \textit{et al.} \cite{Rissanen2021} and De Lucena \textit{et al.} \cite{deLucena2021} did not specify the types of tasks performed, they likely incorporated activities commonly found in a normal daily routine. Another similarity was the settings where the recordings of upper limb activity happened. Most of the studies took place directly within the participants' homes \cite{Saleh2016, Leuenberger2017, Goodwin2021, Lang2021, Pau2021, Rissanen2021, Yang2021, Bandini2022, deLucena2022, Esbjornsson2022, Fortune2022, Kos2007, Waddell2018, Waddell2019, Pohl2022}. The importance of recording at home lies in its ability to capture the performance domain of the ICF, allowing for situations that closely mimic the real-life experiences of individuals with neurological disorders in their daily environments. These home-based settings, though entirely unstructured and requiring a more complex hardware integration, offer a natural and familiar environment for monitoring UE activities. In other studies \cite{Urbin2015, Held2018, Adams2021, deLucena2021, Flury2021}, data collection initially commenced in clinical settings before transitioning to community-based recordings. This approach facilitated a valuable comparison between controlled clinical conditions and real-world home environments.

The overall recording duration and the maximum duration of a single recording varied considerably among studies, reflecting the diverse objectives and protocol designs, as well as the different technological approaches adopted (\textit{i.e.}, battery life is shorter for a camera compared to an accelerometer). Considering the total duration, some studies employed relatively short monitoring periods, such as less than 5 hours \cite{Tsai2020, Bandini2022} and between 5 and 10 hours \cite{Flury2021, Pohl2022, Held2018}. Longer monitoring durations were adopted in other studies, with duration over 2-3 days \cite{Leuenberger2017, Adams2021, Kos2007, Urbin2015, Waddell2018, Waddell2019, Kos2007, Saleh2016, Goodwin2021}. In \cite{Saleh2016}, part of the study was conducted for a two-week period, and a similar duration was achieved in \cite{Pau2021}, while in \cite{deLucena2022}, the study lasted three weeks. Lastly, in \cite{Esbjornsson2022, Fortune2022} authors opted for more extended 30-day monitoring periods. 

Regarding the maximum single duration of a recording session, some studies, such as Adams \textit{et al.} (2021) \cite{Adams2021} and Rissanen \textit{et al.} (2021) \cite{Rissanen2021}, conducted continuous monitoring sessions spanning an impressive 45 and 48 hours, respectively. Similarly, in \cite{ Yang2021} authors recorded data continuously for 48 hours, only during waking hours. In \cite{Lang2021, Leuenberger2017, Kos2007, Urbin2015, Waddell2018, Waddell2019}, the maximum duration for a single session of data recording was 24 hours. De Lucena\textit{ \textit{et al.}} \cite{deLucena2022} recorded data for around 12 hours in a single session. In contrast, shorter periods were reported in \cite{Held2018} (3h) and in \cite{Bandini2022} (1.5h). It is worth noting that for some studies, the maximum duration of a single session was not explicitly specified (\textit{e.g.}, \cite{Esbjornsson2022, Fortune2022}). 

\subsection{Data processing}\label{sec:dataProcessing}
As a consequence of the hardware selection, in the majority of research papers UE performance measures were derived from accelerometry data alone or in conjunction with other signals, including angles \cite{Leuenberger2017}, and angular velocity \cite{Fortune2022}. If considering IMUs as a single-mode sensing technology (IMUs fuse data from 2-3 sources but the resulting outcome is purely kinematic), interestingly, only in two works a multimodal sensing strategy was implemenetd. In \cite{Rissanen2021}, authors used data both from accelerometers and sEMG placed on both forearms of 23 individuals with EPM1 to measure their kinematics and muscular activity; in \cite{deLucena2021,deLucena2022}, authors embedded in their wrist-mount device an IMU and four magnetometers, and asked the participant to wear a magnetic ring on the index finger: in this way, they were able to track both wrist and index finger movements.

The majority of studies opted for an offline data analysis, where data were only recorded and stored during the intervention, while processed only after the monitoring session has concluded \cite{Adams2021, deLucena2021, Flury2021, Goodwin2021, Lang2021, Pau2021, Rissanen2021, Yang2021, Bandini2022, Fortune2022, Held2018, Kos2007, Urbin2015, Waddell2018, Waddell2019}. Interestingly, however, a subset of studies employed real-time analysis, allowing for immediate processing and feedback during monitoring \cite{Saleh2016, deLucena2022, Esbjornsson2022}. Saleh \textit{et al.} \cite{Saleh2016} equipped their glove with an array of light-emitting diodes (LED) to inform the user about the finger flexion ROM. De Lucena \textit{et al.} \cite{deLucena2022} had a screen in their wrist-watch showing the hand movement count and hand use intensity metrics (see below for more information), and an emoji representing their recent performance towards their daily goals. Esbjornsson \textit{et al.} \cite{Esbjornsson2022}, instead, in case of an imbalance in arm movements, made their bracelets vibrate, requiring participants to take an app-based test on their smartphone to detect the onset of a stroke.

Regarding the measures used to quantify UE performance, the most prevalent metrics were activity count and activity duration. Activity count metrics encompass various measures that quantify the number of upper limb actions during daily life, typically expressed as the number of actions per unit of time. Activity duration metrics, on the other hand, quantify the duration of these activities in time units. Other less common measures included the type of activity performed, frequency metrics that described upper limb motion, and traditional biomechanical measurements such as range of motion (ROM) and muscle activity. Table \ref{table_measures} provides an overview of the types of measures used to quantify UE performance at home or in community settings.

\begin{table*}[!htbp]
    \centering
    \begin{adjustbox}{max width=\textwidth}
    \begin{tabular}{cccC{0.28\textwidth}cccc}      
 \textbf{Study} & \textbf{Clinical} & \textbf{Sample} & \textbf{Technology, Sensor Number} & \textbf{Multimodal} & \textbf{Monitoring} & \textbf{Remote} & \textbf{Real-Time} \\
 \textbf{Reference} & \textbf{Condition} & \textbf{Size} & \textbf{and Sensor Location} & \textbf{Sensing} & \textbf{Duration} &  \textbf{Access} & \textbf{Feedback} \\
\midrule
\arrayrulecolor{black!10}
Kos \textit{et al.} 2007 \cite{Kos2007} & MS & 19 & Accelerometer (2x) on wrist and ankle & & $\circ$ 3 days  & & \\ \midrule
Urbin \textit{et al.} 2015 \cite{Urbin2015} & Stroke & 35 & Accelerometers (2x) on wrists & & $\circ$ 22 hours &  & \\ \midrule
\multirow{2}{*}{Saleh \textit{et al.} 2016 \cite{Saleh2016}}  & \multirow{2}{*}{Stroke} & 6 & Bending sensors (2x) on index and &  &  $\circ$ 2 days &  & \multirow{2}{*}{\checkmark} \\ 
 & & 1 & middle fingers & & $\circ$ 2 weeks & &  \\ \midrule
Leuenberger \textit{et al.} 2017 \cite{Leuenberger2017} & Stroke & 10 & IMU (5x) on wrists, shanks and waist & & $\circ$ 2 days & & \\ \midrule
\multirow{2}{*}{Held \textit{et al.} 2018 \cite{Held2018}} & \multirow{2}{*}{Stroke} & \multirow{2}{*}{4} & IMU (14x) full-body suit (1x foot, & & \multirow{2}{*}{9 hours} &  & \\
&&& 2x leg, 3x arm, 2x torso) &&&& \\ \midrule
Waddell \textit{et al.} 2018 \cite{Waddell2018} & \multirow{2}{*}{Stroke} & 64 & \multirow{2}{*}{Accelerometers (2x) on wrists} &  &  2 days &   &  \\ 
Waddell \textit{et al.} 2019 \cite{Waddell2019} & & 29 & & & 5 days & & \\ \midrule
\multirow{2}{*}{Tsai \textit{et al.} 2020 \cite{Tsai2020}} & Stroke & 9 & \multirow{2}{*}{Egocentric camera (1x) on the head}&  &  \multirow{2}{*}{3 hours} &   &  \\ 
 & SCI & 14 & & & & & \\  \midrule
\multirow{2}{*}{Adams \textit{et al.} 2021 \cite{Adams2021}} & \multirow{2}{*}{PD} & \multirow{2}{*}{17} & Accelerometers (5x) on thighs,  & & \multirow{2}{*}{$\circ$ 2 days} & & \\
&&& forearms and trunk &&&& \\  \midrule
Flury \textit{et al.} 2021 \cite{Flury2021} & Stroke & 15 & IMU (6x) on chest, forearms, legs (3x) & & $\circ$ 5 hours &  & \\ \midrule
Goodwin \textit{et al.} 2021 \cite{Goodwin2021} & SCI & 34 & IMU (3x) on chest and upper arms & & $\circ$ 2 days  & & \\ \midrule
Lang \textit{et al.} 2021 \cite{Lang2021} & Stroke & 67 & Accelerometers (2x) on wrists & & 8 days & & \\  \midrule
Pau \textit{et al.} 2021 \cite{Pau2021} & MS & 28 & Accelerometers (2x) on wrists & & $\circ$ 2 weeks & & \\ \midrule
\multirow{2}{*}{Rissanen \textit{et al.} 2021 \cite{Rissanen2021}} & \multirow{2}{*}{EPM1} & \multirow{2}{*}{23} & EMG (2x) and accelerometers (2x) on & \multirow{2}{*}{\checkmark} & \multirow{2}{*}{$\circ$ 2 days} & & \\ 
&&& forearms &&&& \\  \midrule
Yang \textit{et al.} 2021 \cite{Yang2021} & Stroke & 14 & Force Myography (1x) on wrist & & $\circ$ 3 days & & \\ \midrule
Bandini \textit{et al.} 2022 \cite{Bandini2022} & SCI & 20 & Egocentric camera (1x) on the head & &4.5 hours & & \\ \midrule
de Lucena \textit{et al.} 2021 \cite{deLucena2021} & \multirow{2}{*}{Stroke} & 29 & IMU (1x) and magnetometers (4x) on wrist & \multirow{2}{*}{\checkmark} & $\circ$ 6-9 hours & \multirow{2}{*}{\checkmark} & \multirow{2}{*}{\checkmark}\\
de Lucena \textit{et al.} 2022 \cite{deLucena2022} & & 20 & and magnetic ring (1x) on the index finger & & $\circ$ 3 weeks & & \\ \midrule
Esbjornsson \textit{et al.} 2022 \cite{Esbjornsson2022} & Stroke & 30 & Accelerometers (2x) on wrist & & $\circ$  30 days &  \checkmark & \checkmark \\ \midrule
Fortune \textit{et al.} 2022 \cite{Fortune2022} & SCI & 10 & IMU (3x) on chest and upper arms & & $\circ$ 1.67 days & & \\ \midrule
\multirow{2}{*}{Pohl \textit{et al.} 2022 \cite{Pohl2022}} &	\multirow{2}{*}{Stroke} & \multirow{2}{*}{14} & IMU (2x) on wrists and a camera for & & \multirow{2}{*}{$\circ$ 6.5 hours} & & \\ 
&&& activity labelling &&&& \\ 
\arrayrulecolor{black!100}
\midrule
\end{tabular}
\end{adjustbox}
    \vspace{.5em}
    \caption{State of the art of wearable technologies to monitor individuals with neurological impairments in the community.
    IMUs were considered single-mode sensors (kinematic data) despite fusing multiple sources of information (gyroscope, accelerometer, and magnetometer). 
    SCI = Spinal Cord Injury, MS = Multiple Sclerosis, PD = Parkinson's Disease, EPM1 =  Progressive Myoclonic Epilepsy type 1, IMU = Inertial Measurement Unit. $\circ$ = consecutive period. 
    }
    \label{table_sota}
\end{table*}

\textit{Activity count} was typically estimated from acceleration data by calculating the magnitude of the raw acceleration within a specific window length \cite{Leuenberger2017, Pohl2022, Kos2007, Esbjornsson2022}. Consequently, it was essential to establish an optimal threshold for acceleration magnitude to distinguish between functional and non-functional movements. For example, in \cite{Pohl2022} the authors addressed this issue by determining the optimal threshold for both affected and non-affected arms (\textit{i.e.}, maximizing the area under the curve) and subsequently trained a logistic regression classifier to discern functional from non-functional interactions using raw IMU data. Since recordings were often conducted in unconstrained settings, it was necessary to filter out walking periods by detecting lower limb accelerations through shank accelerometers \cite{Leuenberger2017}. The results were typically presented as the average counts per unit of time (\textit{e.g.}, per minute or hour). 

Activity count was also estimated from angles using various methods:
\begin{itemize}
    \item Ratio of movement \cite{Saleh2016}, as movement episodes detected by applying a 2-degree threshold to identify changes in a specific joint angle (in this case, finger flexion). The ratio of counted samples over the entire dataset yielded the ratio of movement values for each finger.
    \item Integral of individuated movement \cite{Saleh2016}, a parameter related to the mean difference in angle between two fingers, representing individual finger movement.
    \item Gross arm movement identification \cite{Leuenberger2017}, based on forearm elevation orientation, with specific criteria for defining gross arm movement.
\end{itemize}

Finally, less common strategies to estimate activity counts included data from magnetic fields \cite{deLucena2021,deLucena2022}, force myography \cite{Yang2021}, and egocentric video \cite{Bandini2022}.

\textit{Activity duration} metrics were commonly computed from accelerometry data, often determined by summing the seconds in which the acceleration magnitude exceeded either zero \cite{Lang2021} or a predefined threshold \cite{Pau2021}. Angular data were also used to calculate the percentage of time spent in different elevation bins \cite{Goodwin2021}, offering insights into humeral elevation angles. Angular velocity data \cite{Flury2021} was used to extract arm activity data, employing the Hilbert transform and a binary threshold to identify active arm periods. The arm activity time was then determined according to this definition during the recording time. Video data could also be used to calculate the average duration of hand-object interaction, providing insights into activity duration \cite{Bandini2022, Tsai2020}.

\textit{Activity type} was determined by some authors who delved beyond quantifying the active time of the upper limbs and considered the type of actions and activities being performed. From accelerometry data, metrics such as bilateral magnitude and magnitude ratio \cite{Lang2021, Pau2021, Urbin2015, Waddell2018, Waddell2019} were used to assess both upper limb use. These metrics offered insights into the intensity and contribution of each limb to daily activity on a second-by-second basis. Mono and bilateral arm use index \cite{Pau2021} quantified the frequency of independent movements in everyday activities. Use ratio \cite{Lang2021, Urbin2015, Waddell2018, Waddell2019} measured the total duration of activity of one limb relative to the other. Density plots graphically represented accelerometry data from both limbs, providing visual insights into movement patterns. In contrast, in \cite{Fortune2022} authors employed machine learning to estimate activity types, using a set of 13 features as predictors in a neural network model.

\textit{Frequency measures} were used to understand the type of motion, particularly in conditions like Parkinsonism, where tremors at specific frequencies are important predictors of treatment efficacy. These measures included the maximum of the acceleration power spectrum, as well as the amplitude and frequency of rhythmic hand motion \cite{Adams2021}.

\textit{Muscle activity} analysis, although less explored in this context, due to the comfort offered by accelerometers and IMU (especially for long recordings), is noteworthy. Rissanen \textit{et al.} \cite{Rissanen2021} employed several measures, including sample kurtosis, correlation dimension, recurrence rate, root-mean-square amplitude, and burst frequency, to assess EMG signal characteristics related to muscle activity.

\subsection{Comparison between performance metrics and clinical assessments}

When evaluating wearable technologies for monitoring UE performance at home or in community settings, validation against clinical assessments is crucial. Indeed, more than half of the studies included a comparison of the collected data with clinical assessment scores. Considering that most of the reviewed studies focused on stroke, the most common clinical evaluations were the Fugl-Meyer Assessment - Upper Extremity subscale (FMA-UE) \cite{Fugl-Meyer1975} and Action Research Arm Test (ARAT) \cite{McDonnel2008}.

Yang et al. \cite{Yang2021} found a decreased functional activity in the affected hand in individuals with lower FMA-UE scores, indicated by a strong negative correlation (Spearman's correlation $\rho$ = -0.70) between FMA-UE scores and the asymmetry index. De Lucena et al. \cite{deLucena2021} reported a positive correlation between FMA-UE scores and hand use intensity (Pearson's correlation \textit{r} = 0.68), with a weaker correlation for upper extremity activity (\textit{r} = 0.42). However, Pohl et al. \cite{Pohl2022} found no significant correlation between acceleration thresholds and FMA-UE scores. Concerning ARAT, Urbin et al. \cite{Urbin2015} observed strong correlations with performance measures like use ratio ($\rho$ = 0.79) and magnitude ratio ($\rho$  = 0.83). Waddel et al. \cite{Waddell2019} and Lang et al. \cite{Lang2021} both found that lower initial ARAT scores correlated with greater potential for improvement.

Studies using the Box and Block Test (BBT) \cite{Leuenberger2017} showed moderate to high correlations with total paretic arm activity counts (\textit{r} = 0.69 to \textit{r} = 0.93). De Lucena et al. \cite{deLucena2021} also found correlations between BBT scores and both hand use intensity (\textit{r} = 0.67) and UE activity intensity (\textit{r} = 0.64). Pau et al. \cite{Pau2021} used the BBT and 9-Hole Peg Test (9HPT) to evaluate MS patients' UE function. BBT scores positively correlated with minutes of use and vector magnitude (VM) counts (\textit{r} = 0.51 and \textit{r} = 0.55), whereas 9HPT scores showed negative correlations with these measures ($\rho$ = -0.56 and $\rho$ = -0.54), where shorter times indicated better performance.

In individuals with cervical SCI, Bandini et al. \cite{Bandini2022} used the Upper Extremity Motor Score (UEMS) and the graded redefined assessment of strength sensibility and prehension (GRASSP) \cite{KalsyRyan2012}. They found moderate-to-strong positive correlations between egocentric measures of hand use and bilateral UEMS. Dominant hand correlations were weak-to-moderate compared to unilateral scores, while non-dominant hand correlations were moderate-to-strong. In PD patients, Adams et al. \cite{Adams2021} found that tremor time correlated with maximal at-rest scores for both OFF and ON states (assessed through the unified Parkinson's disease rating scale -- UPDRS \cite{UPDRS}). The strongest correlation was in the right hand OFF condition (\textit{r} = 0.79). Real-world tremor proportions were strongly correlated with clinical assessments for both hands (\textit{r} = 0.88 for right; \textit{r} = 0.87 for left).

Rissanen et al. \cite{Rissanen2021} assessed myoclonus in EPM1 using the unified myoclonus rating scale (UMRS) \cite{UMRS}. They found that accelerometry measures had stronger correlations with UMRS scores than EMG features. Maximum acceleration power spectrum showed very strong correlations for both arms ($\rho$ = 0.90 for dominant; $\rho$ = 0.80 for non-dominant), whereas EMG features had moderate to strong correlations.

Lastly, Waddell et al. \cite{Waddell2018} assessed UE functions post-stroke with the Motor Activity Log (MAL) \cite{MAL}, a self-report questionnaire. They discovered a moderate association between MAL scores and use ratio at both baseline ($\rho$ = .31) and after intervention ($\rho$ = 0.52).

Table \ref{table_measures} presents an overview of the metrics employed to measure UE performance, along with details on the correlations between at-home measurements and clinical evaluations.

\begin{table*}[!ht]
    \centering
    \begin{adjustbox}{max width=\textwidth}
    \begin{tabular}{ccccC{0.45\textwidth}}      
 \textbf{Study} & \textbf{Clinical} & \textbf{Clinical} &  \textbf{Type of} &  \\
 \textbf{Reference} & \textbf{Condition} & \textbf{Assessment} &  \textbf{Measure} & \textbf{Performance Measure vs Clinical Assessment} \\
\midrule
\arrayrulecolor{black!10}
Kos \textit{et al.} 2007 \cite{Kos2007} & MS & N/A & Activity count & N/A \\ \midrule
Urbin \textit{et al.} 2015 \cite{Urbin2015} & Stroke & ARAT & Activity type &  Increased paretic UE activity with higher ARAT scores \\ \midrule
Saleh \textit{et al.} 2016 \cite{Saleh2016} & Stroke & FMA-UE & \multicolumn{1}{c}{\begin{tabular}[c]{@{}c@{}}Activity count\\ Activity duration\end{tabular}} &  N/A \\ \midrule
Leuenberger \textit{et al.} 2017 \cite{Leuenberger2017} & Stroke & BBT & Activity count &  Increased paretic UE activity and duration with higher BBT scores \\ \midrule
Held \textit{et al.} 2018 \cite{Held2018} & Stroke & \multicolumn{1}{c}{\begin{tabular}[c]{@{}c@{}}FMA-UE\\ ARAT\end{tabular}} & \multicolumn{1}{c}{\begin{tabular}[c]{@{}c@{}}Activity count\\ Activity type\end{tabular}} & N/A \\ \midrule
Waddell \textit{et al.} 2018 \cite{Waddell2018} & Stroke & \multicolumn{1}{c}{\begin{tabular}[c]{@{}c@{}}ARAT\\ MAL\end{tabular}} & Activity type & Increased paretic paretic UE use with higher MAL scores \\ \midrule
Waddell \textit{et al.} 2019 \cite{Waddell2019} & Stroke & ARAT & \multicolumn{1}{c}{\begin{tabular}[c]{@{}c@{}}Activity duration \\ Activity type\end{tabular}} & Higher increase in paretic UE use in individuals with lower ARAT scores at the beginning of the study \\ \midrule
Tsai \textit{et al.} 2020 \cite{Tsai2020} & \multicolumn{1}{c}{\begin{tabular}[c]{@{}c@{}}Stroke\\ SCI\end{tabular}} & N/A & \multicolumn{1}{c}{\begin{tabular}[c]{@{}c@{}}FMA-UE\\
ARAT\\ UEMS\end{tabular}} & N/A \\ \midrule
Adams \textit{et al.} 2021 \cite{Adams2021} & PD & UPDRS & Frequency measures & Increased time spent with tremors with higher at-rest tremor (as per UPDRS)\\ \midrule
Flury \textit{et al.} 2021 \cite{Flury2021} & Stroke & \multicolumn{1}{c}{\begin{tabular}[c]{@{}c@{}}FMA-UE\\ ARAT\\ MAL\end{tabular}} &  Activity duration & N/A \\ \midrule
Goodwin \textit{et al.} 2021 \cite{Goodwin2021} & SCI & DASH & Activity duration &N/A \\ \midrule
Lang \textit{et al.} 2021 \cite{Lang2021} & Stroke & \multicolumn{1}{c}{\begin{tabular}[c]{@{}c@{}}FMA-UE\\ ARAT\end{tabular}} & \multicolumn{1}{c}{\begin{tabular}[c]{@{}c@{}}Activity duration\\ Activity type\end{tabular}} & Faster UE recovery in individuals with higher ARAT  scores at 4 weeks post-stroke \\ \midrule
Pau \textit{et al.} 2021 \cite{Pau2021} & MS & \multicolumn{1}{c}{\begin{tabular}[c]{@{}c@{}}BBT\\ 9-HPT\end{tabular}} & \multicolumn{1}{c}{\begin{tabular}[c]{@{}c@{}}Activity count\\ Activity duration\end{tabular}} & Increased activity count and duration with higher BBT scores. Decreased count and duration with increased 9-HPT scores. \\ \midrule
Rissanen \textit{et al.} 2021 \cite{Rissanen2021} & EPM1 & UMRS & Muscle activity & Increased muscle activity and jerks with higher UEMS scores \\ \midrule
Yang \textit{et al.} 2021 \cite{Yang2021} & Stroke & \multicolumn{1}{c}{\begin{tabular}[c]{@{}c@{}}FMA-UE\\ REACH\end{tabular}} & Activity count & Higher asymmetrical UE use with lower FMA-UE scores \\ \midrule
Bandini \textit{et al.} 2022 \cite{Bandini2022} & SCI & \multicolumn{1}{c}{\begin{tabular}[c]{@{}c@{}}UEMS\\ GRASSP\end{tabular}} & \multicolumn{1}{c}{\begin{tabular}[c]{@{}c@{}}Activity count\\ Activity duration\end{tabular}} & Higher activity time and higher number of interactions in individuals with more residual hand function \\ \midrule
de Lucena \textit{et al.} 2021 \cite{deLucena2021} & Stroke & \multicolumn{1}{c}{\begin{tabular}[c]{@{}c@{}}UEMS\\ BBT\end{tabular}} & Activity count & Higher paretic UE use with higher FMA-UE \\ \midrule
de Lucena \textit{et al.} 2022 \cite{deLucena2022} & Stroke & \multicolumn{1}{c}{\begin{tabular}[c]{@{}c@{}}FMA-UE\\ ARAT\\ BBT\end{tabular}} & Activity count & N/A \\ \midrule
Esbjornsson \textit{et al.} 2022 \cite{Esbjornsson2022} & Stroke & N/A & Activity count & N/A \\ \midrule
Fortune \textit{et al.} 2022 \cite{Fortune2022} & SCI & N/A & Activity type & N/A \\ \midrule
Pohl \textit{et al.} 2022 \cite{Pohl2022} &	Stroke & \multicolumn{1}{c}{\begin{tabular}[c]{@{}c@{}}FMA-UE\\ ARAT\end{tabular}} & Activity count & Acceleration thresholds for functional UE use did not correlate with FMA-UE scores\\ 
\arrayrulecolor{black!100}
\midrule
\end{tabular}
\end{adjustbox}
    \vspace{.5em}
    \caption{Overview of the metrics employed to measure UE performance, along with details on the correlations between at-home measurements and clinical evaluations. N/A = information not available or not reported in the article; ARAT = Action Research Arm Test; BBT = Box and Block Test; DASH = Disability of the Arm, Shoulder, and Hand; EPM1 = Progressive Myoclonic Epilepsy type 1; FMA-UE = Fugl-Meyer Assessment - Upper Extremity subscale; GRASSP = Graded Redefined Assessment of Strength Sensibility and Prehension; MAL = Motor Activity Log; MS = Multiple Sclerosis; PD = Parkinson's Disease; REACH = Rating of Everyday Arm-use in the Community and Home; SCI = Spinal Cord Injury; UEMS = Upper Extremity Motor Score; UMRS = Unified Myoclonus Rating Scale; UPDRS = Unified Parkinson's Disease Rating Scale; 9-HPT = 9-Hole Peg Test.}
    \label{table_measures}
\end{table*}

\section{Discussion}

In this review, we explored the latest advancements in wearable technologies designed for real-life and unstructured monitoring of UE function in individuals with neurological disorders. Our specific objectives encompassed several crucial aspects, such as identifying and categorizing the most commonly used technologies for monitoring UE function during daily activities, as well as gaining insights into the prevailing methods for data processing and measurement, and investigating the types of protocols developed for conducting such studies.

\subsection{Stroke as the main monitored condition}
As in the case of other wearable technologies \cite{Proietti2022}, stroke survivors comprised the majority of the study population (64\%). This prevalence can likely be attributed to the fact that stroke ranks as one of the leading global causes of both mortality and disability, with UE impairments as a common consequence. The significance of UE recovery for stroke survivors should not be understated, as regaining UE functionality ranks among their highest rehabilitation priorities \cite{Purton2023, Rudberg20202}. This underscores the research community's keen interest in monitoring individuals at home to gain insights into their UE usage during daily life. Such knowledge serves as a foundation for developing therapies aimed at optimizing UE functions at home, with the ultimate goal of enhancing individuals' independence.

Notably, some of the reviewed papers extended their focus beyond chronic cases, including acute and sub-acute stroke survivors in their investigations. For example, Waddell \textit{et al.} \cite{Waddell2019} demonstrated that sensor-measured UL performance improves in the first 12 weeks post-stroke, proving with data from unsupervised conditions the well-known spontaneous capacity of the body to improve after stroke \cite{Ramsey2017}. Lang \textit{et al.} \cite{Lang2021}, instead, showed how UL performance in daily life reached a plateau only 3-6 weeks post-stroke, thus often before neurological impairments and functional capacity started to stabilize. The availability of data early after a stroke is particularly valuable as it addresses a critical phase in patients' recovery. Once patients are discharged from the hospital, the progress of their rehabilitation may become challenging to track. Having a means of observing and understanding their performance upon returning to the community assumes paramount importance. This monitoring is essential for further enhancing their recovery, especially during outpatient care, and ensures that the gains made in the clinical setting continue to progress effectively in the real-world context \cite{Toh2023, Gopal2022, Sica2021, Kim2022, Alexander2021}.

In most of the reviewed studies (12/21, 57\%), authors compared data collected in the community through wearable monitoring technologies with clinical assessments performed in clinics, often involving correlation analyses. This validation process is crucial as it forms the basis for developing novel outcome measures of UE performance that reflect real-life hand use. In many cases, it was shown that improved UE function according to clinical scales corresponded to increased UE usage at home, such as heightened engagement in daily activities. While this comparison approach has been widely adopted (see Table \ref{table_measures}), it represents only the initial phase in developing new outcome measures for hand function. To achieve further validation, future studies must incorporate these monitoring systems and metrics into longitudinal investigations, to evaluate the effectiveness of various rehabilitation interventions over time.

\subsection{Monitoring kinematics only is not enough}
The majority of the studies predominantly relied on IMUs and accelerometers. These choices were motivated by user-friendliness, extended recording capabilities spanning hours or days, and the devices' lightweight and affordability. Moreover, these technologies are very mature as they are widely used in many commercial applications beyond patient monitoring (\textit{e.g.}, modern smartphones embed IMUs). Such attributes make IMUs and accelerometers the natural choices for studies that require non-intrusiveness and portability to collect ecological data. A noteworthy observation from Figure \ref{piecharts} is that the studies utilizing accelerometers tended to have the largest participant populations \cite{Waddell2018, Lang2021}, and reached a favorable balance between expanding the number of recruited participants and maintaining extended recording durations \cite{Esbjornsson2022, Pau2021}.
In this review, we excluded studies using smartphone-embedded IMUs as a source of information, given that they are not properly wearable devices. However, they can provide a very user-friendly and affordable sensing solution to easily assess a large number of individuals (see \textit{e.g.}, Pratap \textit{et al.} who monitored almost 500 MS individuals with this strategy \cite{Pratap2020}).

It is important to note that IMUs and accelerometers primarily offer global kinematic information, which can subsequently be processed to calculate activity metrics like counts and durations. In fact, given the scope of this review -- where included papers monitored patients during daily life to quantify performance -- the use of activity metrics is pretty intuitive, as the performance domain directly relates to how individuals carry out activities in their typical environment. Yet, for a more in-depth analysis, particularly to understand specific types of grasps and the contextual nuances of UE functions, IMUs and accelerometers alone may prove insufficient.

In response to this limitation, some studies have introduced video-based approaches, capitalizing on the advantages of recording richer information about the surrounding environment \cite{Tsai2020, Bandini2022}. This includes details such as the manipulation area, objects, and the broader context, which aids in deciphering the functional aspects of hand-object interactions and the specific activities being performed. However, it is worth noting that wearable cameras may not be optimal for extended recordings due to comfort issues and battery runtime \cite{Bandini2021}. Alternatively, some studies have explored the use of magnetic sensors to gather additional information about finger movements \cite{deLucena2021, deLucena2022}, while others have delved into EMG recordings \cite{Rissanen2021}. Magnetic sensors offer enhanced detail but may introduce artifacts when interacting with metal objects, while EMG, although valuable for capturing low-level muscle activation, was not extensively examined for deciphering grasp patterns, as already seen in other studies \cite{Toro-Ossaba2022}. 

While this review discussed several wearable technologies (e.g., accelerometers, IMUs, cameras, etc.), there exist other approaches to monitoring using different strategies that were not covered by this work (e.g., tattoo sensors \cite{Ferrari2018} ). Challenges such as robustness -- in terms of sensor materials and data cleanliness (e.g., absence of drift and low noise), especially when used unsupervised in unconstrained environments -- along with long-term usability, still present barriers to their adoption for monitoring UE functions at home. These issues hinder their transition into marketable products. Despite these obstacles, these advanced technologies hold significant potential to positively impact the field, and further validation studies are anticipated in the near future. 

In the meantime, adopting multi-modal approaches that combine the strengths of different sensors could offer a more comprehensive understanding. However, the majority of the reviewed studies predominantly relied on single-mode technologies, with only Rissanen \textit{et al.} \cite{Rissanen2021} implementing a true multimodal sensing strategy by using data from both accelerometers and sEMG placed on the forearms of 23 individuals with EPM1. The scarcity of multi-modal strategies may be attributed to the numerous challenges that need to be addressed when approaching multi-modal monitoring. Firstly, data processing from multiple sensors is generally more complex. Integrating data streams with different characteristics, such as sampling rates and noise levels, requires sophistication in calibrating the sensors, fusing, and synchronizing the data, thus increasing the computational burden and complicating analysis. Moreover, managing a richer dataset presents significant storage, transfer, and processing challenges.
From a hardware standpoint, implementing multi-modal sensing often results in larger, heavier, and more expensive setups. Integrating multiple sensors into wearable devices affects their form factor and portability, making them less practical for daily use.
Finally, the limited number of studies exploring multi-modal sensing in UL activity monitoring may discourage researchers from pursuing this approach. Single-sensor technologies are more established and easier to implement, potentially overshadowing the benefits of multi-modal sensing.

\subsection{Short, offline recording, with poor feedback to the patients}
Concerning data processing, the majority of analyses were conducted offline. This approach may be suitable for tracking recovery progress in neurorehabilitation, where changes and improvements may be appreciated over weeks or months. Reviewing wearable technologies through the lens of the performance domain can be advantageous for telerehabilitation techniques. Such approaches enable a direct evaluation of the interventions' influence on patients' performance. Additionally, they offer benefits like motivating rehabilitation efforts and fostering a sense of accomplishment or satisfaction in patients, who can witness the therapy's impact on their daily lives.

Nonetheless, real-time data analysis offers distinct advantages, particularly for providing immediate feedback to patients during telemonitoring and telerehabilitation \cite{Hribernik2022}. It is important to highlight a few articles that were excluded \cite{Sanders2020, Dodakian2017, Toh2023b}. While these studies concentrated on telerehabilitation and did not include monitoring ADLs, they utilized compelling wearable methods to study the UEs at home in people with neurological conditions.
Besides, real-time processing, with only the processed data transmitted, may even help address privacy concerns, which is a critical issue associated with technologies like those based on cameras \cite{Bandini2021}.

When it comes to activity metrics, the most prevalent measures focus on quantifying the duration and frequency of activities. However, in a rehabilitation context, understanding not only the quantity but also the context of these measures is crucial for identifying specific challenges patients encounter during different activities. Machine learning techniques, as proposed by Fortune \textit{et al.} \cite{Fortune2022}, have the potential to recognize the types of activities being performed. This information, combined with the quantity of hand use, is essential for gaining a comprehensive understanding of UE use at home, thus quantifying the performance domain.

Furthermore, as shown in Figure \ref{piecharts}, the duration of recording sessions is typically limited to a few hours in the majority of cases. While some of these studies often explored the feasibility of using technology at home, it is essential to extend the recording duration to encompass a broader range of daily living activities typically performed by participants during their daily routines. This ensures that the recorded behavior in the home environment closely reflects the participants' everyday activities. This is particularly important in light of the findings by Waddell \textit{et al.} \cite{Waddell2018} showing how self-reports UL performance are neither consistent nor accurate with sensor-based use assessment. A potential explanation for the limited duration of the recordings may be attributed to wearability and comfort concerns. However, as demonstrated in Section \ref{sec:stateoftheart}, almost none of the studies considered these factors in their analyses. This presents a significant issue because the ability to extend recording duration, aside from factors like battery size and data storage, is closely linked to the device's acceptance by participants. This acceptance, in turn, is correlated with factors such as comfort, appearance, and weight. Adding analysis of these parameters, for instance via surveys or mixed method studies \cite{Saleh2016, Bandini2021}, is crucial and should be targeted in future works. One potential cause of this gap is that most of the studies utilized commercially available devices (such as the Motionlogger, the Link, or the GT3X accelerometers by \textit{Actigraph}, the BioStampRC IMUs by \textit{MC10}, the Emerald or the Opal IMUs by \textit{APDM}, the Xsens IMUs suit by \textit{Movella}, or the Hero 4 and 5 cameras by \textit{GoPro}), potentially considering comfort outside their scope and relying on the manufacturer. However, this approach may be limited, and highlighting comfort and wearability in monitoring studies remains fundamental to understanding our current stance on this specific topic. 

\begin{figure}
    \centering
    \includegraphics[width = 0.5\textwidth]{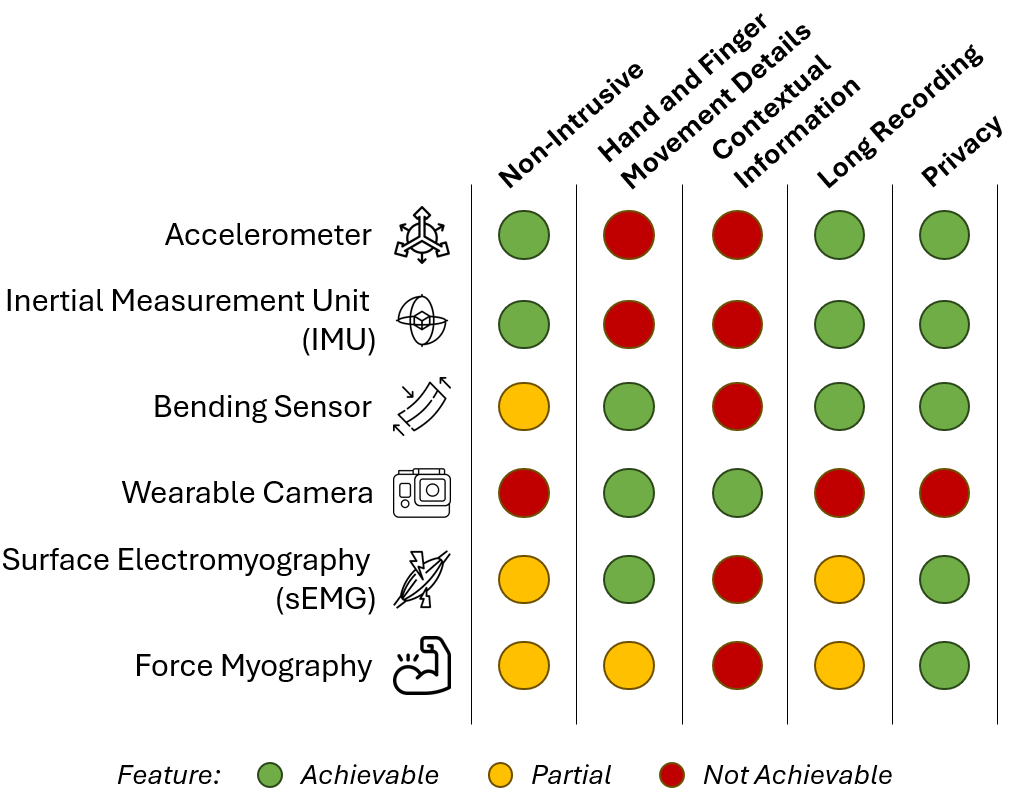}
    \caption{Comparison of Wearable Technologies for Monitoring UE Function at Home in Individuals with Neurological Impairments.}
    \label{fig:comparison}
\end{figure}

\subsection{Final remarks and future directions}
The results of this literature review underscore the growing significance of monitoring UE performance remotely in the neurorehabilitation field, with the majority of the research conducted in this domain emerging over the past 15 years, and a notable surge in the last three years. This trend can be attributed to the increasing accessibility and affordability of off-the-shelf technologies, facilitating the recording of extended periods of unconstrained activities in the community. The potential unlocked by this capability is indeed promising, offering the opportunity to tailor rehabilitation strategies to better suit individuals' specific needs within their daily lives.

However, it is crucial for researchers to consider several key aspects while pursuing these opportunities. Firstly, ensuring device usability is essential, as these technologies are intended for use by non-expert individuals in their home environments. This user-friendliness is vital to guarantee the quality of the recorded data. Secondly, it is essential to consider the potential challenges associated with attaching sensors to their arms and hands, as this may impact how they perform daily activities. Thirdly, privacy concerns may arise from monitoring individuals in their homes, especially in the case of video monitoring, and this must be addressed.  
In considering the future directions of wearable technology for monitoring UL function, several key considerations emerge, each essential for advancing the field and addressing the diverse needs of users.

\textit{1) Non-intrusive and Lightweight Design.} It is paramount to ensure the wearables to be non-intrusive and lightweight, as they must seamlessly integrate into users' daily lives, minimizing discomfort and inconvenience. Examples of such devices include wristbands or sensors embedded within clothing. Prioritizing comfort and ease of use facilitates continuous wear over extended periods, enabling comprehensive monitoring and analysis of UL performance.

\textit{2) Detailed Capture of Hand and Finger Movements.} Effective differentiation of hand and finger movements is also fundamental for gesture recognition and task assessment. Wearables should be capable of capturing detailed information about these movements with high accuracy and precision. Finger-worn devices \cite{deLucena2021, deLucena2022} or sensorized gloves \cite{Oess2012, Saleh2016} may provide granular data for nuanced analysis. However, they are limited in detecting contextual information, which instead may be provided by video-based approaches \cite{Tsai2020, Bandini2022}.

\textit{3) Integration of Contextual Information} Complementary to the previous point, contextual information enriches the understanding of UE function, enhancing the usefulness of wearable devices. Multimodal strategies could enable the capture of contextual cues essential for interpreting movements within their environmental context. Advanced algorithms and machine learning techniques for action and activity recognition could also facilitate the extraction of meaningful insights from rich datasets collected at home.

\textit{4) Long-term Monitoring Capabilities.} The ability to perform long recordings spanning several hours is imperative for comprehensive assessment and monitoring of UE function. Challenges such as power consumption, data storage, and user comfort must be addressed to facilitate continuous monitoring over extended periods. Efficient power management strategies, data compression techniques, and cloud-based storage solutions offer avenues for overcoming these challenges, enabling sustained monitoring without compromising device usability or performance.

\textit{5) Privacy Considerations.} As wearable technology continues to proliferate, ensuring the privacy and security of user data is paramount. Wearable devices must be designed with robust privacy safeguards, including data encryption, user-controlled data access, and anonymization techniques. Adherence to relevant privacy regulations, such as the General Data Protection Regulation (GDPR) and Health Insurance Portability and Accountability Act (HIPAA), is essential to instill trust and confidence among users and stakeholders.

By looking at the above review, it becomes evident that a singular technology that fulfills all these requirements is currently nonexistent (see Fig. \ref{fig:comparison}). This underscores the increasing need for a multimodal approach to effectively capture how individuals function in their home environments. \hl{Moreover, the availability of long recordings makes this field ideal for implementing machine learning and deep learning approaches. Some reviewed studies already use these methods to recognize activity types or extract hand location and contact state information from raw videos} \cite{Fortune2022, Bandini2022, Tsai2023} \hl{. Regardless of the hardware and data collected, machine learning and deep learning will be essential for interpreting the vast amounts of data collected at home and summarizing it into simple measures for tracking upper extremity function progress. With new extended and multi-modal datasets, AI will become increasingly important, addressing key requirements such as recognizing contextual information and summarizing long-term recordings, ultimately providing interpretable information to therapists and clinicians to enhance patient care.}

\end{document}